\documentclass[runningheads]{llncs}
\usepackage{subcaption}

\usepackage{graphicx}
\usepackage{booktabs}
\begin{document}
\title{An Empirical Study on Learning Paths and Gender Dynamics in Scrum Master Roles}
%
%
\author{Manuela Petrescu\inst{1}\orcidID{0000-0002-9537-1466} \and
Paul Razvan Petrescu\inst{1}\orcidID{0009-0004-4215-1525} }
\authorrunning{Petrescu et al.}
%
\institute{
Department of Computer Science, Babes Bolyai University, Cluj-Napoca, Romania\\
\email{manuela.petrescu@ubbcluj.ro,rpetrescu@gmail.com}}
\maketitle              
\begin{abstract}
\textbf{Context}: Agile development methodology has been widely adopted by industry and the demand for experienced professionals in Agile-related roles is persistently high.
\textbf{Objectives}: We focus on the learning path for a Scrum Master role in multicultural software companies and investigate the role in relation to team size, together with the learning process for a career path, and how companies monitor soft skills development.
\textbf{Method}: We conducted our study in two phases, two qualitative surveys (interview studies) and performed a qualitative and quantitative data analysis of the results. 
\textbf{Conclusions}: Our results identified that the need for a Scrum Master (SM) depends on the size of the team, with our study indicating a six-member limit. There is no overall standardized process for soft skills learning or metrics to measure progress. Some companies measure soft skills based on feedback received from the client or from the team, and other companies are taking both types of feedback into consideration. Many learning initiatives, especially on soft skills for an SM role, were based on employees' actions.

\keywords{Scrum Master learning \and career path \and Agile \and software.}
\end{abstract}
\section{Introduction}
\label{SIntro}

The Software Product Development Lifecycle (SDLC) has experienced an ongoing evolution over time to address the requirements of contemporary projects. 
There are different methodologies based on Agile framework such as Scrum, KanBan, Extreme Programming. Scrum framework was widely adopted, offering benefits such as team planning, continuous iterations, and consistent testing \cite{b33}. According to the Scrum Guide (2020), a Scrum team consists of three key roles: the Scrum Master, the Product Owner, and the developers.
The role of the Scrum Master includes acting as a facilitator who has the ability to motivate and understand the team.


In response to a growing demand in the market, certain companies have successfully facilitated the inclusion of people lacking technical experience in the engineering industry, particularly in roles related to computer science such as Scrum Master (SM) or Business Analyst (BA)  \cite{b4}. 
There are studies on how to create systems using Scrum \cite{b32}, others are related to the importance of the Scrum Master position,  \cite{b34,b36}, gender diversity in this role \cite{b35,b18}, soft skills for the role of SM are investigated in \cite{b34,b19} but studies that address the learning path for the position of Scrum Master are relatively few. This is the main reason we focus on the Scrum Master role. We focus on gender, as the study \cite{b20} has indicated that there is a tendency for men to occupy more technical roles within the field of computer science, while women tend to be more prevalent in positions that require greater emphasis on soft skills and fewer technical competencies. 

Skills can be categorized into hard/technical - specialized knowledge and soft skills. Soft skills refer to a set of personal attributes, behaviours, and characteristics of an individual. This research paper examines the learning path for a Scrum Master from an empirical perspective, completing studies by \cite{b10} and \cite{b28} which have explored the crucial hard and soft skills required in Scrum.
The following articles are related to our work in terms of methodology, using interviews \cite{b10}, \cite{b24} and focus groups \cite{b10} to acquire qualitative data, and in terms of results when verifying the necessary skills \cite{b9}, \cite{b10} for the Scrum Master position and exploring a potential career path in this field.

This study serves as an extended continuation of the research presented in \cite{b26}, which focused on identifying the necessary competencies, responsibilities, and gender balance for the role of SM. The main research questions in the preliminary paper were related to finding the technical and soft skills essentially linked to the SM role and to finding out if 'the Scrum Master role was doubled by managerial or technical duties'. 
The objective of this study was to determine the learning path for a Scrum Master role within the context of Agile development methodology, as well as to examine career-related facts of women in relation to Scrum Master positions. To achieve this goal, we have formulated the objectives \textbf{1:} Understand and identify the relationship between the size of the team and the need for an SM in a team.
\textbf{2:} Comprehend the learning process for a career path for women in SM positions.
\textbf{3:} Examine how companies monitor the skills development associated with Scrum Master positions.

\section {Study Design}
\label{sec:interviewDesign}
This section describes the design of the study and information on data collection. The structure of this study conforms to established community standards for Qualitative Surveys \cite{b27}. For this study, we used qualitative approaches, more specifically, to conduct two phases of interviews using Teams to record them. 
Having interviews enabled us to follow a structure but at the same time let the respondents answer in free text about their own opinion around the subject. We intended to position the need for a SM role in different teams, depending on their size, and to discover the learning path for this position in different software companies. The current paper is an extension of \cite{b26}, where we explored the skills necessary to perform the SM role.

For each phase, we had different sets of questions for the interviews, some of the questions were specifically designed to focus on women's roles. The first set consisted of questions related to the company's specificity, personal information such as current role, overall IT experience, company type: [multinational, outsourcing], size of the team, and also questions related to Agile, questions 1 to 5. The interviewees were asked to describe the Agile process used, to specify learning processes related to the SM roles. The second set was specific for women who perform Scrum Master roles in IT development teams. The second set of questions was divided into two sections: The first section was used for group selection and data processing, including details such as the interviewer's name and interview status. The second section focused on how women perform the Scrum Master role within specific project contexts. The second set of questions (except for the specificity questions) is listed in Table \ref{tab:questions}.
\vspace{-20pt}
\begin{table}[h!]
  \caption{Interview Questions}
  \label{tab:questionnarie}
  {\small{
  \begin{tabular}{p{12cm}}
\toprule
    What type of background do you have (technical/nontechnical). \\ 
    Please describe previous positions.\\
    Please describe the career path to the SM role. \\ 
    Did you have SM specific preparation (workshops, courses, conferences, tutorials)? \\
    Did you have mentors for the SM career path? \\
    Please describe the support received from your organization. \\
   \bottomrule
\end{tabular}
}}
\vspace{-20pt}
\label{tab:questions}
\end{table}

\subsection{Participants}
\label{sec:participants}

For each phase, we had different participant selection criteria. In the first phase, we had individual discussions with 32 people from 14 companies, to be included in our study they had to have significant prior experience in Agile methodology and/or in fulfilling the role of an SM, to hold for at least 5 years a leadership or managerial position that involves leading, supervising, or overseeing a development team, to guarantee representation across various company sizes. We tried to promote gender diversity (when feasible, we prioritized interviews with women over men, since the majority of eligible candidates for interviews were men), we had 14 women and 18 men. In the second phase, for the second round of interviews, we searched for women who have experience and worked in SM positions; the selection criteria did not take into account the size of the company or the size of the team.
The average experience for the set of participants was 13.64 years in the computer science field. In terms of company size, the participants worked in six small companies and eight large companies (company size was determined using the criteria outlined in EUROSTAT (2022)). There were no biases in establishing the participant set, as we initiated the study participation invitation to all our collaborators and encouraged them to extend invitations to other persons who met the eligibility criteria.


\subsection{Methodology}
\label{sec:methodology} 

\textbf{Data collection} process followed the standard procedures outlined in \cite{b27} for the interviews, and for the focus group. The interviews were conducted online and recorded with the consent of the interviewee. We paid attention to ethical concerns and communicated the study objectives, used anonymization, and obtained consent for the final transcript. Two project members were responsible for conducting the interviews and writing the interview transcripts. An independent third project member, uninvolved in both the interview and transcript creation processes, reviewed the transcript by cross-referencing it with the recorded interview. 
During the first phase (32 interviews), more than 33 hours were dedicated to conducting interviews and an additional 52 hours were invested in reviewing the data and generating transcripts for the first set of interviews; at the end of that period,  the discussion in the focus group with SM professionals lasted more than three hours. The second phase consisted of a series of 12 interviews with women who performed Scrum master roles in IT teams. For the second phase, we dedicated around 8 hours to conducting interviews and another 17 hours to review data and generate the transcripts. We have adopted a reflexive methodology for data analysis, taking into account the subsequent steps based on \cite{b14}. 
Consequently, for free-text responses, we selected thematic analysis as recommended by \cite{b14}, described in \cite{b26} and previously used in other research studies in software engineering \cite{b5,b25}.

The focus group was organized according to existing standard procedures and good practices recommended by \cite{b27,b31}. The focus group meeting was led by two of the co-authors, one acting as a moderator and another as an assistant. The focus group participants, already familiar with the study's topic from interviews, shifted the three-hour discussion from project-specific details to a broader exploration of their experiences and learning in the SM role.

\section{Empirical investigation into Scrum Master learning path - Results}
\label{empirical-investigation}

\subsection{RQ1: What is the need for a Scrum Master role depending on the size of the team?}

To find the answer to RQ1, we analysed the data from the first set in interviews to see if there is a correlation between the size of the team and the need for a SM role and the gender of the person who performs this role. We discussed the responses received for the questions: \textit{''Please specify the team size'', ''Specify if there was a person having SM position'', ''Please specify the gender''}.

We found that the largest software development team had 70 members (a multicultural team of Romanians, Indians, and others), and the smallest had only four members. More than half of the teams had less than 10 members. 
With one exception, all teams having more than six members had a person in the SM position, even if, for some teams, the person having this role performed other tasks and/or roles. 
Teams that worked using Agile and did not have an SM role had 6 or less than 6 members. In conclusion, the need for an SM correlates with the size of the software team; in our case, the size limit was six. 
In our study, the role of SM was performed by 29.41\% women and 57.14\% men. So, even if women are a good fit based on the skills required for an SM role \cite{b21}, in our study, most of the time, this role was carried out by men.
\\

\subsection{RQ2. How is the learning process for a career path for women in Scrum Master roles?}

To find the answer to this question, we focused on the women in our study who were performing in Scrum Master roles. The distribution of our respondents from technical background point of view is: 7 have technical background in terms of traditional academic degree, 3 come from business/economics background, one from engineering field (other than computer), and one from European studies background. Although all of them acquired some kind of experience in technical concepts through previous positions or internships as students. In some cases (3 persons) they possess relatively simple programming skills (testing, business analyst). In one case, the respondent declared that she considers business skills essential and more important than technology (programming). 

\begin{figure}[!htbp]
\centering
\includegraphics[width=0.4\textwidth]{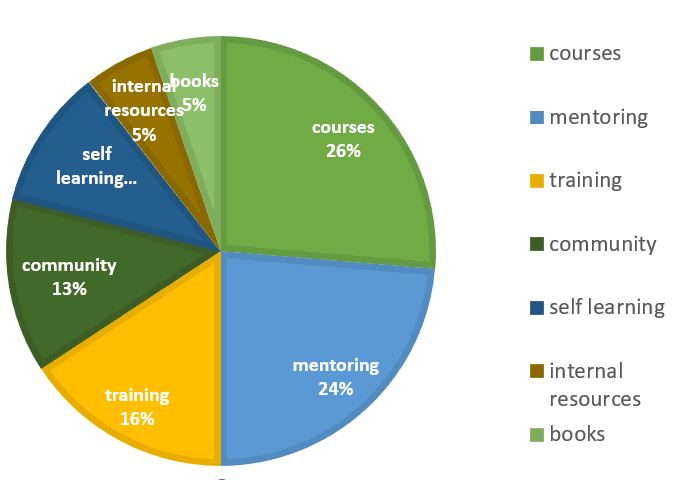}
\includegraphics[width=0.55\textwidth]{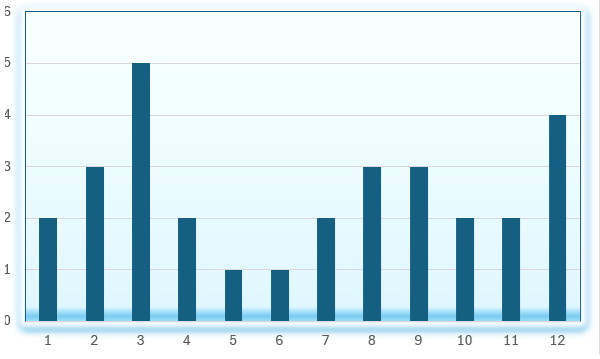}
\caption{(a) Map of learning forms and (b) Support offered by company: number of distinct supporting activities for each respondent} 
\label{fig_rq2}
\end{figure}

Different forms of learning have been identified during interviews, such as courses, finalized with/without certification (either on site or online), community learning (forum, local meetups) and self-learning. A complete map of these forms can be seen in Figure \ref{fig_rq2}(a).

We might notice the diverse forms of learning, in which courses, including online courses, and mentoring represent 50\% of them. It is also worth remarking that 53\% of learning methods, mentoring, trainings, and community learning are interactive, which is important as experience and communication are some essential characteristics of the SM role. We found that the learning forms are very different and that the abilities for such a role are gathered from different learning experiences, as described by the respondents: \textit{''we had a mentor/coach for one to one discussion; we had daily syncs to make sure we get results'', ''It helped me to coach other colleagues about event adoption, other Scrum issues'', ''Community learning: Reddit, local meetups''}. As a notice, we did not correct, modify, or edit the participant's responses, and we cited them as they were, so possible English language issues or short forms of the verbs might appear. Obtaining a certification might not be of high importance, as only 4 out of 11 interviewed persons mention that after course completion they went to the certification process. 

The support offered by the organization was also very diverse, from very low offered (\textit{ 'No support from the company except the opportunity to follow this path”, “No official program for the learning and development part (except Udemy and PMP)”, “time to learn”}), to continuous support in different forms: courses, certification, funding to follow courses, training programs, mentoring, shadowing programs. Without differentiating between different supporting actions, if we simply count these actions, the distribution is shown in Figure \ref{fig_rq2} (b), with the remark that company support can, in most cases, be more consistent.

Conclusion: The interactive learning forms are prevailing, providing means to gain facilitation abilities; Support of organization, even present in more cases, can be more consistent and can be provided in more forms, thus up-skilling own human resources to Scrum Master role, rather than searching for external ones and many learning initiatives came from employees.

\subsection{RQ3.  How do companies monitor the skills development for a Scrum Master role?}

According to the results of \cite{b26}, soft skills are a must for a Scrum Master role, as having technical skills is a plus. Because of this, we focused on the responses for soft skills improvement. We noticed that most of the software companies have defined metrics for technical skills measurement such as number of bugs resolved, number of implemented features, team velocity; they formalized technical evaluations for teams and for individuals, also using code review, certifications, etc. We believed there was little innovation in measuring technical skills, so we did not pay attention to this topic. However, companies did not define metrics to measure soft skills, and thus the subject is worthy of investigation.

According to our data, soft skills are measured using feedback: in 53.12\% of the responses. Sometimes feedback comes from the client, sometimes from the team: \textit{''Soft skills in one to one to manager and based on the client’s feedback'', ''Peer review: questionnaire completed by everyone who worked with''}. The next largest category of answers (31.25\%) mentioned that nothing is standardized for the evaluation of soft skills, it is more an informal process: \textit{''For soft skills we have intuitive measures, nothing standardized'', ''We don’t have KPIs for soft skills; for soft skills we do it intuitively and punctual''}. Evaluations are performed twice a year (18.75\%), once a year (15.65\%) or every month (9.37\%). Other responses stated that developing soft skills is \textit{''each one's responsibility''}, and other stated that certifications are taken into account. There were two answers that stated how soft skills development is encouraged, but nothing is mandatory: \textit{''There is no formal tracking for soft skills development, but there is encouragement within the company to participate in internal soft skills training programs, especially for junior to mid levels (as there are multiple in each bracket), and there are more advanced programs for leadership and management designed for mid and higher levels'', ''Additionally, within the team we also encourage initiatives in communication with the client, preparing demo presentations, providing internal trainings and constant feedback for these activities on the project''}. We also got answers such as \textit{''not measured''} and one answer stating \textit{''No idea''}.

Conclusion: There is no overall standardized process for soft skills; for some companies, everything is based on feedback received from the client or from the developing team; other companies are taking both types of feedback into consideration. None of the companies have defined metrics for "how well someone communicates, chairs a meeting, negotiates or solves a conflict", everything is in one big bucket, general feedback from the client and/or team.

\section{Discussion}
\label{sec:results}
Soft skills have a longer life span, but it is harder to quantity them and define specific metrics. Technical skills acquirement and progress are easier to measure compared to soft skills, and software companies invest in up-skilling their employees technical skills by different methods: courses, pair-programming, mentoring, etc. 

As indicated in the RQ1 findings, the necessity of a Scrum Master (SM) depends on the size of the team, our study indicating a threshold of six members. Although women possess the necessary skills for the Scrum Master (SM) role \cite{b21}, our study found that men predominantly filled this position. This suggests potential gender imbalances in Agile leadership, highlighting the need for more research on organizational biases, cultural factors, and opportunities to encourage more women in SM roles. The process of acquiring Scrum Master skills varies widely and is influenced by the company culture, personal motivation, and available resources. Some organizations emphasize structured training programs, while others rely on mentorship and hands-on experience. 
Interactive learning methods, such as workshops, simulations, and peer discussions, are particularly effective in developing facilitation abilities.
Companies do not have a standardized method to evaluate soft skills, employees often take the initiative in their learning journey, seeking training, certifications, and practical experiences to improve their skills.

There is no universal standard for assessing soft skills in the SM role. Some companies rely solely on feedback from clients and teams, while others consider both perspectives. However, none have established concrete metrics to evaluate communication, facilitation of meetings, negotiation, or conflict resolution. The absence of standardized evaluation methods makes it challenging to measure and improve these critical skills systematically. A more structured approach, including defined benchmarks and assessment tools, could help organizations better develop and refine these competencies. Establishing clear evaluation criteria would enhance leadership effectiveness and contribute to a more consistent and efficient Agile environment.

\section{Threats to validity}
\label{sec:threats}

As an empirical investigation, the study has taken measures to address threats to validity. The authors adhered to established guidelines \cite{b27} and took into account the following factors: construct validity, internal validity, and external validity. 
We took steps to anonymize the data by removing identification information about the individuals and companies involved. Construct validity refers to the pertinence and coherence of the interview. Before conducting the interviews, two persons with expertise as SM and Product Manager validated the proposed set of questions. Their feedback was used to calibrate the questions.

Internal validity refers to the various factors that exert an influence on the results of a study. To address the potential risks posed by the participant pool, we implemented a strategy aimed at diversifying the composition by deliberately selecting individuals who occupied various roles and positions and ensuring a balanced gender representation, taking into account the size of the company.

External validity: We examined how the results of this study can be generalized to a wider range of companies and how can they be replicated in different settings and different types of projects \cite{b37}. The key to answer to this question lies in the selection of the participants (from 14 multinational or outsourcing companies) plus 7 additional companies from the focus group. Because of this, we consider the results can be generalized to Romanian IT companies. Generalizing the results for IT companies in general can only be done with some caution, since Romania exports around 80\% of the ITC products (Information Technology and Cybersecurity) according to the CEIC report \cite{b38}. However, Romanian companies can have some specificity related to internal processes and learning methodologies.

\section{Conclusions and future work}
\label{sec:conclusion}

Throughout this study, we have investigated the learning path for a Scrum Master to thrive in the technological industry, to comprehend the support offered by software companies, and to analyze women's' training for a Scrum Master position. In order to do this, we have organized two sets of interviews and a focus group.
The data collected showed that the acquisition of Scrum Master skills varies according to the company culture, personal motivation, and availability of resources. Although organizational support is prevalent, there is potential to improve its consistency and diversity, as many learning initiatives were created by employees. Our findings indicate that software companies monitor technical skills development, as soft skills for an SM role are difficult to monitor and evaluate, as most companies based their evaluation on the feedback received, without having defined any metrics. 

In conclusion, it is our contention that this study has significance in determining the learning path in software companies for the position of Scrum Master. In future work, we intend to extend our investigation on the learning process to other roles within Agile teams.

%
%
%
%

\end{document}